\documentclass[aps,prc,showpacs,twocolumn]{revtex4}

\usepackage{graphicx}
\usepackage{dcolumn}

\newcommand{\ba}{\begin{eqnarray}}
\newcommand{\ea}{\end{eqnarray}}

\begin{document}
\title{Partial dynamical symmetry as a selection criterion for many-body 
interactions}

\author{A.~Leviatan$^1$,
J.E.~Garc\'\i a-Ramos$^2$,
and P.~Van Isacker$^3$}
\affiliation{
$^1$Racah Institute of Physics,
The Hebrew University, Jerusalem 91904, Israel}
\affiliation{
$^2$Department of Applied Physics, University of Huelva,
21071 Huelva, Spain}
\affiliation{
$^3$Grand Acc\'el\'erateur National d'Ions Lourds,
CEA/DSM--CNRS/IN2P3, B.P.~55027, F-14076 Caen Cedex 5, France}
\date{\today}

\begin{abstract}
We propose the use of  partial dynamical symmetry (PDS)
as a selection criterion  for higher-order terms in situations
when a prescribed symmetry is obeyed by some states
and is strongly broken in others.
The procedure is demonstrated in a first systematic classification
of many-body interactions with SU(3) PDS
that can improve the description of deformed nuclei. 
As an example, the triaxial features of the nucleus $^{156}$Gd are analyzed.
\end{abstract}

\maketitle

Many-body forces play an important role in quantum many-body 
systems~\cite{3body}. 
They appear either at a fundamental level or as effective interactions 
which arise due to 
restriction of degrees of freedom and truncation of model spaces. 
A known example is the structure of light nuclei, where 
two-nucleon interactions are insufficient to achieve an 
accurate description and higher-order interactions
between the nucleons must be included~\cite{Argonne}.
Given the difficulty in constraining
the nature of such higher-order terms from experiments,
one is faced with the problem of their determination.
One way, currently the subject of active research~\cite{machleidt11},
is to determine them from chiral
effective field theory applied to 
quantum chromodynamics. 
This establishes a hierarchy of inter-nucleon interactions 
according to their order.
In light-medium nuclei, these interactions serve as input for ab-initio 
methods ({\it e.g.}, the no-core shell model (NCSM)~\cite{NCSM}) 
to generate, by means of similarity transformations, 
A-body effective Hamiltonians in computational tractable model spaces.

The situation is more complex in heavy nuclei, 
where ab-initio methods are limited by 
the enormous increase in size of the model spaces required to 
accommodate correlated collective motion 
of many nucleons. One possible approach 
to circumvent this problem, is to augment the NCSM method through 
a symplectic symmetry-adapted choice of basis~\cite{Sp-NCSM}. 
A second approach is to employ 
energy density functionals and incorporate beyond mean-field 
effects by mapping to collective Hamiltonians~\cite{EDF-IBM}, 
{\it e.g.}, the interacting boson model (IBM)~\cite{Iachello87a}. 
In both approaches the Hilbert spaces are based on particular 
dynamical algebras which lead to a dramatic reduction of the basis 
dimension. Nevertheless, even with such simplification, 
the number of possible interactions in the effective Hamiltonians 
grows rapidly with their order,
and a selection criterion is called for. 
In this Rapid Communication, we suggest a method to 
select possible higher-order terms 
which is based on the idea of partial dynamical symmetry (PDS).

The concept of PDS~\cite{Leviatan11} 
is a generalization of that of a dynamical symmetry (DS)~\cite{Iachello06}
where the conditions of the latter
(solvability of the complete spectrum,
existence of exact quantum numbers for all eigenstates,
and pre-determined structure of the eigenfunctions)
are relaxed and apply to only {\em part}
of the eigenstates and/or of the quantum numbers. 
PDSs have been identified in various dynamical 
systems involving bosons and fermions 
(for a review, see Ref.~\cite{Leviatan11}). 
They play a role in diverse phenomena including 
nuclear and molecular spectroscopy~\cite{Leviatan96,fermionPDS,Ping97}, 
quantum phase transitions~\cite{lev07} and mixed regular and chaotic 
dynamics~\cite{WAL93}. 
Here we consider the SU(3) symmetry in view of its significance 
for deformed nuclei, as recognized in the Elliott 
and symplectic shell models~\cite{Elliott58,SSM} 
and the IBM. 
We use the mathematical algorithm to construct, order by order,
all possible interactions with a given PDS~\cite{Alhassid92,GarciaRamos09},
apply it to the SU(3) limit of the IBM,
and illustrate with a concrete example
how the PDS and data constrain the 
form and strength of higher-order interactions.
\begin{table*}
\caption{\label{t_tensors}
Normalized two- and three-boson SU(3) tensors.}
\begin{ruledtabular}
\begin{tabular}{ccccl}
$n$&$(\lambda,\mu)$&$\tilde{\chi}$&$\ell$&
$\hat B^\dag_{[n](\lambda,\mu)\tilde{\chi}\ell m}$\\[4pt]
\hline
2& (4,0)& 0& 0& $\sqrt{\frac{5}{18}}(s^\dag)^2
                           +\sqrt{\frac{2}{9}}(d^\dag d^\dag)^{(0)}_0$\\[2pt]
2& (4,0)& 0& 2& $\sqrt{\frac{7}{9}}s^\dag d^\dag_m
                           -\sqrt{\frac{1}{9}}(d^\dag d^\dag)^{(2)}_m$\\[2pt]
2& (4,0)& 0& 4& $\sqrt{\frac{1}{2}}(d^\dag d^\dag)^{(4)}_m$\\[2pt]
2& (0,2)& 0& 0& $P^\dag_0\equiv
                             -\sqrt{\frac{2}{9}}(s^\dag)^2
                           +\sqrt{\frac{5}{18}}(d^\dag d^\dag)^{(0)}_0$\\[2pt]
2& (0,2)& 0& 2& $P^\dag_{2m}\equiv
                             \sqrt{\frac{2}{9}}s^\dag d^\dag_m
                           +\sqrt{\frac{7}{18}}(d^\dag d^\dag)^{(2)}_m$\\[2pt]
3& (6,0)& 0& 0& $\sqrt{\frac{7}{162}}(s^\dag)^3
             +\sqrt{\frac{14}{45}}s^\dag(d^\dag d^\dag)^{(0)}_0
             -\sqrt{\frac{8}{405}}((d^\dag d^\dag)^{(2)}d^\dag)^{(0)}_0$\\[2pt]
3& (6,0)& 0& 2& $\sqrt{\frac{7}{30}}(s^\dag)^2d^\dag_m
              -\sqrt{\frac{2}{15}}s^\dag(d^\dag d^\dag)^{(2)}_m
              +\sqrt{\frac{2}{21}}((d^\dag d^\dag)^{(0)}d^\dag)^{(2)}_m$\\[2pt]
3& (6,0)& 0& 4& $\sqrt{\frac{11}{30}}s^\dag(d^\dag d^\dag)^{(4)}_m
            -\sqrt{\frac{14}{165}}((d^\dag d^\dag)^{(2)}d^\dag)^{(4)}_m$\\[2pt]
3& (6,0)& 0& 6& 
            $\sqrt{\frac{1}{6}}((d^\dag d^\dag)^{(4)}d^\dag)^{(6)}_m$\\[2pt]
3& (2,2)& 0& 0& $W^\dag_0\equiv
            -\sqrt{\frac{1}{9}}(s^\dag)^3
            +\sqrt{\frac{1}{20}}s^\dag(d^\dag d^\dag)^{(0)}_0
            -\sqrt{\frac{7}{180}}((d^\dag d^\dag)^{(2)}d^\dag)^{(0)}_0$\\[2pt]
3& (2,2)& 0& 2& $V^\dag_{2m}\equiv
             -\sqrt{\frac{14}{65}}(s^\dag)^2d^\dag_m
             -\sqrt{\frac{1}{130}}s^\dag(d^\dag d^\dag)^{(2)}_m
             +\sqrt{\frac{18}{91}}((d^\dag d^\dag)^{(0)}d^\dag)^{(2)}_m$\\[2pt]
3& (2,2)& 2& 2& $W^\dag_{2m}\equiv
              \sqrt{\frac{2}{39}}(s^\dag)^2d^\dag_m
             +\sqrt{\frac{14}{39}}s^\dag(d^\dag d^\dag)^{(2)}_m
             +\sqrt{\frac{5}{78}}((d^\dag d^\dag)^{(0)}d^\dag)^{(2)}_m$\\[2pt]
3& (2,2)& 2& 3& $W^\dag_{3m}\equiv
             \sqrt{\frac{7}{30}}((d^\dag d^\dag)^{(2)}d^\dag)^{(3)}_m$\\[2pt]
3& (2,2)& 2& 4& $W^\dag_{4m}\equiv
              \sqrt{\frac{2}{15}}s^\dag(d^\dag d^\dag)^{(4)}_m
             +\sqrt{\frac{7}{30}}((d^\dag d^\dag)^{(2)}d^\dag)^{(4)}_m$\\[2pt]
3& (0,0)& 0& 0& $\Lambda^\dag\equiv
             -\sqrt{\frac{1}{81}}(s^\dag)^3
             +\sqrt{\frac{5}{36}}s^\dag(d^\dag d^\dag)^{(0)}_0
             +\sqrt{\frac{35}{324}}((d^\dag d^\dag)^{(2)}d^\dag)^{(0)}_0$\\
\end{tabular}
\end{ruledtabular}
\end{table*}

The IBM describes low-energy collective states of the nucleus 
in terms of $N$ monopole $(s)$ and 
quadrupole $(d)$ bosons representing pairs of nucleons. 
The dynamical algebra is U(6) with generators in terms of which
operators of all physical observables can be written.
The classification of states in the SU(3) limit is~\cite{Arima78}
\begin{equation}
\begin{array}{ccccccc}
{\rm U}(6)&\supset&{\rm SU}(3)&
\supset&{\rm SO}(3)&\supset&{\rm SO}(2)\\
\downarrow&&\downarrow&&\downarrow&&
\downarrow\\[0mm]
[N]&&(\lambda,\mu)&K&L&&M
\end{array},
\label{e_clas}
\end{equation}
where underneath each algebra
the associated labels are given 
($K$ is a multiplicity label needed in the 
${\rm SU(3)}\supset {\rm SO(3)}$ reduction). 
These define the Elliott basis~\cite{Elliott58}, 
$\vert [N](\lambda,\mu)KLM\rangle$, 
from which the Vergados basis~\cite{Vergados68}, 
$\vert [N](\lambda,\mu)\tilde{\chi}LM\rangle$,
is obtained by a standard orthogonalization procedure. 
The classification~(\ref{e_clas})
assumes a symmetric U(6) irreducible representation (irrep) $[N]$
which is appropriate for the IBM. Apart from terms involving the 
conserved total boson number operator $\hat{N}$, 
a rotational-invariant Hamiltonian with SU(3) DS has the form
\begin{equation}
\hat H_{\rm DS}=
\alpha_1\hat C_2[{\rm SU(3)}]+
\alpha_2\hat C_2[{\rm SO(3)}]+
\alpha_3\hat C_3[{\rm SU(3)}] ~,
\label{e_ds}
\end{equation}
where $\hat C_n[G]$ is the $n^{\rm th}$ order Casimir operator
of the Lie algebra $G$
and $\alpha_i$ are coefficients.
This form exhausts all independent Casimir operators of SU(3) and SO(3),
that is, any other commuting operator
can be written as a function of those appearing in Eq.~(\ref{e_ds}).
$\hat{H}_{DS}$ is completely solvable with eigenenergies
\ba
E_{\rm DS} = \alpha_{1}f_{2}(\lambda,\mu) + \alpha_{2}L(L+1) 
+\alpha_{3}f_{3}(\lambda,\mu) ~,
\label{ene_ds}
\ea
where $f_{2}(\lambda,\mu)\!=\!\lambda^2\!+\!(\lambda\!+\!\mu)(\mu\!+\!3)$ 
and $f_{3}(\lambda,\mu)\!=\!(\lambda\!-\!\mu)(2\lambda\!+\!\mu\!+\!3)
(\lambda\!+\!2\mu\!+\!3)$. The spectrum~resembles 
that of a quadrupole axially-deformed rotor with eigenstates 
arranged in SU(3) multiplets and $K$ corresponds geometrically 
to the projection of the angular momentum on the symmetry axis. 
The Hamiltonian $\hat H_{\rm DS}$ is genuinely many-body
(with interactions that are up to third order in the bosons). 
Its applicability is limited, however, since 
only three independent operators exist, and states 
in different $K$-bands with the same $(\lambda,\mu)L$ 
are degenerate. 
Flexibility can be considerably increased
by introducing interactions with PDS.
The method to construct such interactions
is based on an expansion of the Hamiltonian in terms of tensors 
which annihilate prescribed set of states~\cite{Alhassid92,GarciaRamos09}. 
In the present study, the tensors involve
$n$-boson creation and annihilation operators
with definite character under the SU(3) chain~(\ref{e_clas}),
\ba
\hat B^\dag_{[n](\lambda,\mu)\tilde{\chi}\ell m},\;\;
\tilde B_{[n^5](\mu,\lambda)\tilde{\chi}\ell m}\equiv
(-)^{m}
(\hat B^\dag_{[n](\lambda,\mu)\tilde{\chi}\ell,-m})^\dag.
\;\;
\ea
The SU(3) tensor operators for $n$=2 and 3 
are given in Table~\ref{t_tensors}.
Of particular interest are the operators with $(\lambda,\mu)\neq(2n,0)$
because the corresponding annihilation operators
yield zero when acting on the ground-band members 
$|[N](2N,0)K\!=\!0,LM\rangle$ (and possibly other states).
Interactions involving these operators
can be added to the Hamiltonian~(\ref{e_ds})
without destroying solvability of part of its spectrum.
Two such operators, $P^\dag_0$ and $P^\dag_{2m}$, exist for $n$=2,
and allow the construction of an IBM Hamiltonian with up to 
two-boson interactions
that have a solvable ground band $|[N](2N,0)K\!=\!0,LM\rangle$
and a solvable $\gamma$ band $|[N](2N-4,2)K\!=\!2,LM\rangle$.
A two-body Hamiltonian with SU(3) PDS can be applied to $^{168}$Er
and the excellent SU(3) description of the energies and E2 properties
of these bands can be retained while lifting the degeneracy of 
the $\beta$ and $\gamma$ bands~\cite{Leviatan96}.
\begin{table}
\caption{\label{t_number}
Number of interactions in the IBM.}
\begin{ruledtabular}
\begin{tabular}{cccc}
Order&\multicolumn{3}{c}{Number of interactions\footnotemark}\\
\cline{2-4}
& General & SU(3) DS & SU(3) PDS \\
\hline
$1$ & $2\mapsto1$ & $1\mapsto0$ & $1\mapsto0$\\
$2$ & $7\mapsto5$ & $3\mapsto2$ & $4\mapsto3$\\
$3$ & $17\mapsto10$ & $4\mapsto1$ & $10\mapsto6\phantom{0}$\\
\hline
$1+2+3$ & $26\mapsto16$ & $8\mapsto3$ & $15\mapsto9\phantom{0}$
\end{tabular}
\end{ruledtabular}
\footnotetext[1]{On the left of $\mapsto$
is the number of interactions of a given order;
this reduces to the number on the right of $\mapsto$
if one is only interested in excitation energies in a single nucleus.}
\end{table} 
\begin{figure*}
\begin{center}
\leavevmode
\includegraphics[width=0.7\linewidth]{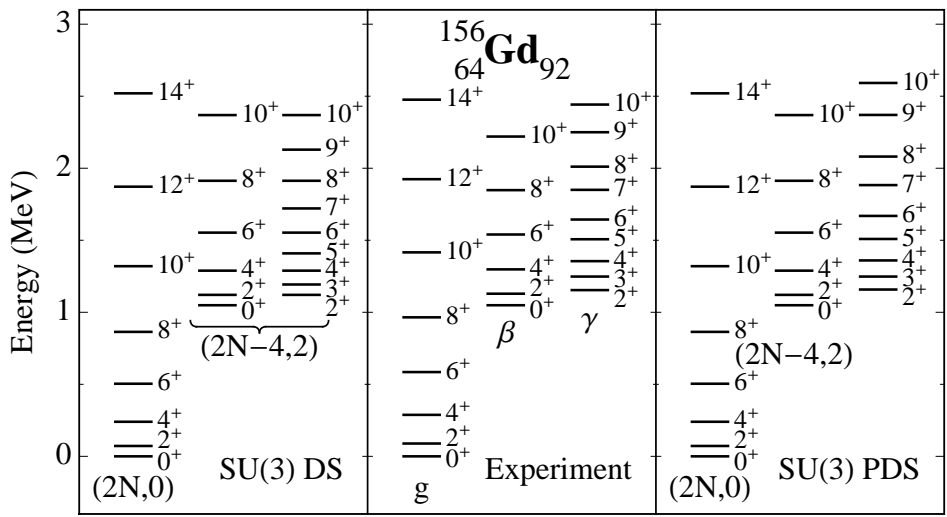}
\caption{
Observed spectrum of $^{156}$Gd~\cite{Reich03}
compared with the calculated spectra
of $\hat H_{\rm DS}$~(\ref{e_ds}) with SU(3) DS
and of $\hat H'_{\rm PDS}$~(\ref{e_pdsp}) with SU(3) PDS 
for $N=12$ and parameters given in the text.
Bands are labeled with the SU(3) quantum numbers $(\lambda,\mu)$.}
\label{f_gd156e}
\end{center}
\end{figure*}

The most general $(2+3)$-body Hamiltonian 
with SU(3) PDS can be written in terms of the operators
given in Table~\ref{t_tensors},
\begin{eqnarray}
\hat H_{\rm PDS}&=&
h_2P^\dag_2\!\cdot\!\tilde P_2+h_0P^\dag_0 P_0+
g_4W_4^\dag\!\cdot\!\tilde W_4+g_3W_3^\dag\!\cdot\!\tilde W_3
\nonumber\\&&+
g_2^aV_2^\dag\!\cdot\!\tilde V_2+
g_2^bW_2^\dag\!\cdot\!\tilde W_2+
g_2^c(V_2^\dag\!\cdot\!\tilde W_2+W_2^\dag\!\cdot\!\tilde V_2)
\nonumber\\&&+
g_0^a\Lambda^\dag\Lambda+
g_0^bW_0^\dag W_0+
g_0^c(\Lambda^\dag W_0+W_0^\dag\Lambda),
\label{e_pds}
\end{eqnarray}
with $2+8$ interactions strengths $h_i$ and $g_i^x$.
Terms involving the operator $\hat C_2[{\rm SO}(3)]\equiv\hat L^2$
can be added to this Hamiltonian, as is done in Eq.~(\ref{e_ds}).
To illustrate the increase in flexibility of a Hamiltonian with SU(3) PDS,
we list in Table~\ref{t_number}
the number of interactions under the different scenarios.
Up to third order, a general rotationally invariant Hamiltonian
has 26 independent interactions,
decreasing to 16
if one is only interested in excitation energies in a single nucleus.
(This excludes terms involving $\hat{N}$).
A Hamiltonian with SU(3) DS has, up to third order,
8 independent terms but 5 of them
($\hat N$, $\hat N^2$, $\hat N^3$, $\hat N\hat L^2$, and 
$\hat N\hat C_2[{\rm SU}(3)]$)
are constant in a single nucleus
or can be absorbed in an interaction of lower order,
leaving only the 3 genuinely independent terms shown in Eq.~(\ref{e_ds}).
The corresponding numbers for a Hamiltonian with SU(3) PDS are 15 and 9.
The latter number agrees with the 10 terms in the Hamiltonian~(\ref{e_pds})
which lacks $\hat L^2$ but includes the combinations
$\hat N P^\dag_2\!\cdot\!\tilde P_2$ and $\hat N P^\dag_0 P_0$.
We conclude from Table~\ref{t_number}
that more than half of all possible interactions in the IBM
have in fact an SU(3) PDS.

Several SU(3)-preserving interactions
are contained in the expression~(\ref{e_pds}). 
Specifically, 
$\hat{\theta}_2\equiv 
2\hat{N}(2\hat{N}\!+\!3) - \hat{C}_{2}$ corresponds to 
$h_0\!=\!h_2\!=\!18$; 
$(\hat{N}\!-\!2)\hat{\theta}_2$: $g_0^a\!=\!54$, 
$g_0^b\!=\!g_2^b\!=\!g_2^a\!=\!g_3\!=\!g_4\!=\!30$; 
$\hat{C}_{3} +
(2\hat{N}\!+\!3)[3\hat{\theta}_2\!-\!2\hat{N}(4\hat{N}\!+\!3)]$: 
$g_0^a\!=\!648$ and 
$\hat\Omega-(4\hat N+3)\hat L^2$: $h_2\!=\!-108$, 
$g_0^a\!=\!9g_0^b\!=\!-3g_0^c\!=\!270$, 
$g_2^a/5\!=\!g_2^b/21\!=\!g_2^c/\sqrt{105}\!=\!24/13$, 
$g_4\!=\!-120$. 
The three terms involving $\hat{C}_{n}[{\rm SU(3)}]$ are 
included in $\hat{H}_{\rm DS}$~(\ref{e_ds}). The (integrity basis) term 
$\hat{\Omega}=-4\sqrt{3}\hat Q\!\cdot\!(\hat L\times\hat L)^{(2)}$
is composed of SU(3) generators, hence is diagonal in $(\lambda,\mu)$, 
but breaks the $K$-degeneracy of the exact DS. 
Its impact on nuclear spectroscopy 
has been well studied in the symplectic shell 
model and the IBM~\cite{Draayer84,Berghe85,Bonatsos88}. 
The PDS notion goes a step further by allowing SU(3) mixing in most 
(but not all) of the eigenstates of the Hamiltonian.

As noted, in general, $\hat{H}_{\rm PDS}$~(\ref{e_pds}) does not preserve 
SU(3) yet, by construction, 
for {\em any} choice of parameters 
the ground-band members $|[N](2N,0)K\!=\!0,LM\rangle$ are solvable. 
For specific choices, additional solvable states are obtained.
In particular, by choosing only the $h_0,\,g_0^a,\,g_0^b,\,g_0^c$ 
terms and $\hat{H}_{\rm DS}$~(\ref{e_ds}), 
the states $|[N](2N-4,2k)K\!=\!2k,LM\rangle$, 
$k\!=\!1,2,\dots$(among which the $\gamma$-band members with $k\!=\!1$)
remain solvable with energies $E_{\rm DS}$~(\ref{ene_ds}). 
This case therefore has the same solvable states
as the two-body Hamiltonian with SU(3) PDS considered 
in Ref.~\cite{Leviatan96};
the additional three-body terms lead
to a different mixing of the non-solvable states.

\begin{figure}
\begin{center}
\leavevmode
\includegraphics[width=\linewidth]{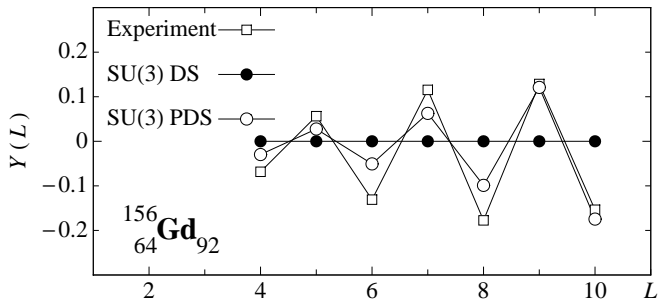}
\caption{
Observed and calculated (SU(3) DS and PDS) 
odd-even staggering
of the $\gamma$ band in $^{156}$Gd.}
\label{f_gd156s}
\end{center}
\end{figure}

Another class of Hamiltonians with SU(3) PDS exists
which has solvable $\beta$-band members 
$|[N](2N-4,2)K\!=\!0,LM\rangle$ 
with energies $E_{\rm DS}$ (\ref{ene_ds}). 
This follows from the structure of the relevant Hamiltonian, 
\begin{equation}
\hat H'_{\rm PDS}=
\hat H_{\rm DS}+
\eta_2W_2^\dag\!\cdot\!\tilde W_2+
\eta_3W_3^\dag\!\cdot\!\tilde W_3,
\label{e_pdsp}
\end{equation}
and the fact that $W_{2m}$ and $W_{3m}$
annihilate the intrinsic state of the $\beta$ band,
$|\beta\rangle\propto
(\sqrt{2}P_0^\dag-P_{20}^\dag)
(s^\dag+\sqrt{2}d_0^\dag)^{N-2}|{\rm 0}\rangle$. 
The property of solvability of the ground and $\beta$ 
bands can be exploited in the following way.
The Hamiltonian $\hat H_{\rm DS}$~(\ref{e_ds}) 
has a rotor spectrum with characteristic $L(L+1)$ splitting for all bands.
Deviations from this pattern are often observed for the $\gamma$ band of 
deformed nuclei and are indicative of $\gamma$-soft or triaxial 
behavior~\cite{Zamfir91}. 
We illustrate the procedure with an application to $^{156}$Gd.
The parameters $\alpha_1=-7.6$~keV, $\alpha_2=12.0$~keV, and $\alpha_3=0$
in $\hat H_{\rm DS}$
are fixed from the excitation energy of the $\beta$-band head
and the moments of inertia of the ground and $\beta$ bands.
This completely determines the SU(3) DS spectrum,
shown on the left of Fig.~\ref{f_gd156e},
which is characterized by degenerate $\beta$ and $\gamma$ bands.
In the observed spectrum these bands are not degenerate
and, more importantly, the $\gamma$-band energies
display an odd-even staggering.
This effect can be visualized by plotting the quantity~\cite{Casten91}
\begin{equation}
Y(L)=
\frac{2L-1}{L}\times
\frac{E(L)-E(L-1)}{E(L)-E(L-2)}-1,
\label{e_stag}
\end{equation}
where $E(L)$ is the excitation energy
of a $\gamma$-band level with angular momentum $L$.
For a rotor this quantity is flat, $Y(L)=0$,
as illustrated in Fig.~\ref{f_gd156s} with the SU(3) DS calculation.
The data, however, show considerable odd-even staggering
which can be well described by a combination
of three-body interactions with $\eta_2=-18.1$~keV and $\eta_3=46.2$~keV.
The calculated staggering increases with $L$
which agrees with the experiment up to $L=10$.
For $L\!>\!10$ the observed staggering changes character, 
a phenomena requiring higher angular momentum pairs, 
which are beyond the 
scope of the standard $(s,d)$ IBM description. The two interactions 
$W_2^\dag\!\cdot\!\tilde W_2$ and $W_3^\dag\!\cdot\!\tilde W_3$
induce a mixing of the $\gamma$ band with higher-lying excited bands.
Other approaches advocating the coupling of the $\gamma$ band to the 
$\beta$ band~\cite{Bonatsos88} or to the ground band~\cite{Minkov00}
fail to describe the odd-even staggering in $^{156}$Gd.
For the PDS calculation, the wave functions of the states in 
the $\gamma$ band involve $15\%$ SU(3) admixtures into the dominant 
$(2N-4,2)$ component. 
Higher bands exhibit larger SU(3) mixing 
and their wave functions are spread over 
many SU(3) irreps, as shown for the $K=0_3$ band in Fig.~3. 
This complex SU(3) decomposition is 
in marked contrast to the SU(3)-purity of the ground ($K=0_1$) and 
$\beta$ ($K=0_2$) bands. 
Such strong symmetry-breaking cannot be treated in perturbation theory.

Previous studies of triaxiality in the 
IBM framework have employed only the cubic $\eta_3$ term 
of Eq.~(\ref{e_pdsp})~\cite{Heyde84,Casten85}. 
The current work hints that both $\eta_2$ and $\eta_3$ terms are 
necessary for an accurate description of odd-even staggering in 
deformed nuclei. This highlights the capacity of the 
PDS approach to identify novel relevant terms of a given order. 
We emphasize that the PDS results for the $\gamma$ band 
are obtained without altering the good agreement for the ground and 
$\beta$ bands, already achieved with the SU(3) DS calculation. 
This is further illustrated with the E2 transitions in $^{156}$Gd.
The observed $B$(E2) values between ground, $\beta$, and $\gamma$ bands 
are shown in Table~\ref{t_be2}
and compared to the results of the SU(3) DS and PDS calculations.
The effective boson charge $e_{\rm b}=0.166$~$eb$ in the electric 
quadrupole operator
$e_{\rm b}[ s^\dag\tilde d +d^\dag s +\chi(d^\dag\tilde d)^{(2)}]$
and the value $\chi=-0.168$
are fitted to the $B({\rm E2};2^+_1\rightarrow0^+_1)$
and $B({\rm E2};2^+_\beta\rightarrow4^+_1)$ values.
The E2 transitions between ground and $\beta$ bands
can be calculated analytically~\cite{Isacker83},
and remain valid in SU(3) PDS.
Transitions involving $\gamma$-band members
are different in SU(3) DS and PDS, 
and are computed numerically for the latter. 
It is seen from Table~\ref{t_be2} that the mixing of the $\gamma$ band 
with higher-lying excited bands improves the agreement with the data 
in most cases.
\begin{figure}
\begin{center}
\leavevmode
\rotatebox{270}{\includegraphics[height=6.9cm]{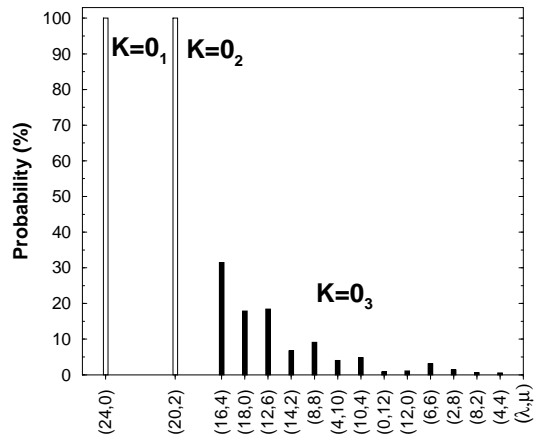}}
\caption{
SU(3) decomposition of wave functions of $L\!=\!0$ states 
in the $K=0_1,\,0_2,\,0_3$ bands for the PDS calculation.}
\label{f_su3decomp}
\end{center}
\end{figure}
\begin{table}
\caption{\label{t_be2}
Observed and calculated $B$(E2) values in $^{156}$Gd.
The parameters of the E2 operator are given in the text.}
\begin{ruledtabular}
\begin{tabular}{rlll}
Transition&\multicolumn{3}{c}
{$B({\rm E2};L^\pi_i\rightarrow L^\pi_f)$~($e^2b^2$)}\\
\cline{2-4}
& Expt~\cite{Reich03}& SU(3) DS & SU(3) PDS \\
\hline
$  2^+_1\rightarrow0^+_1$& 0.933~{\sl 25}  &  0.933 & 0.933  \\
$  4^+_1\rightarrow2^+_1$& 1.312~{\sl 25}  & 1.313 & 1.313  \\
$  6^+_1\rightarrow4^+_1$& 1.472~{\sl 40}  & 1.405 & 1.405  \\
$  8^+_1\rightarrow6^+_1$& 1.596~{\sl 85}  & 1.409 & 1.409 \\
$10^+_1\rightarrow8^+_1$& 1.566~{\sl 70}  & 1.364 & 1.364 \\
$2^+_\beta\rightarrow0^+_\beta$& 0.26~{\sl 11}  & 0.679 & 0.679 \\
$4^+_\beta\rightarrow2^+_\beta$& 1.40~{\sl 75}  & 0.951 &  0.951\\
$0^+_\beta\rightarrow2^+_1$& 0.04~{\sl 2}  & 0.034 & 0.034 \\
$2^+_\beta\rightarrow0^+_1$& 0.0031~{\sl 3}  & 0.0055 & 0.0055 \\
$2^+_\beta\rightarrow2^+_1$& 0.0165~{\sl 15}  & 0.0084 & 0.0084 \\
$2^+_\beta\rightarrow4^+_1$& 0.0204~{\sl 20}  & 0.020 & 0.020 \\
$4^+_\beta\rightarrow2^+_1$& 0.0065~{\sl 35}  & 0.0067 &  0.0067\\
$4^+_\beta\rightarrow4^+_1$&~~~ ---  & 0.0067 & 0.0067 \\
$4^+_\beta\rightarrow6^+_1$& 0.0105~{\sl 55} & 0.021 & 0.021 \\
$2^+_\gamma\rightarrow0^+_1$& 0.0233~{\sl 8}  & 0.035 & 0.030 \\
$2^+_\gamma\rightarrow2^+_1$& 0.0361~{\sl 12}  & 0.056 & 0.048 \\
$2^+_\gamma\rightarrow4^+_1$& 0.0038~{\sl 2}  & 0.0037 & 0.0031 \\
$3^+_\gamma\rightarrow2^+_1$& 0.0364~{\sl 70}  & 0.062 & 0.053 \\
$3^+_\gamma\rightarrow4^+_1$& 0.0254~{\sl 50}  & 0.032 & 0.028 \\
$4^+_\gamma\rightarrow2^+_1$& 0.0090~{\sl 25}  & 0.017 & 0.015 \\
$4^+_\gamma\rightarrow4^+_1$& 0.050~{\sl 15}  & 0.067 & 0.057 \\
$4^+_\gamma\rightarrow6^+_1$&~~~ ---  & 0.0089 & 0.0076 \\
$4^+_\gamma\rightarrow2^+_\beta$& 0.0214~{\sl 80}  & 0.0033 & 0.0096
\end{tabular}
\end{ruledtabular}
\end{table}

In summary, we have identified several classes of $(2+3)$-body 
IBM Hamiltonians with SU(3) PDS, and obtained an improved 
description of signature splitting in the $\gamma$ band of $^{156}$Gd. 
The analysis serves to highlight the merits gained by
using the notion of 
PDS as a tool for selecting 
higher-order terms in systems where 
a prescribed symmetry is not obeyed uniformly. 
On one hand, the PDS approach allows more flexibility 
by relaxing the constraints of an exact DS. On the other hand, 
the PDS picks particular symmetry-breaking terms which do not destroy 
results previously 
obtained with a DS for a segment of the spectrum. The PDS construction 
is implemented order by order, yet the scheme is non-perturbative 
in the sense that the non-solvable states 
experience strong symmetry-breaking. 
These virtues can be exploited in attempts to extend the ab-initio and 
beyond-mean-field methods to heavy nuclei. 
The present work motivates and sets the stage for 
further exploring 
the impact of PDS with higher-order terms 
on the dynamics in quantum many-body systems.
 
This work was supported in part (A.L.) by the Israel Science Foundation
and in part (J.E.G.R.) by the Spanish Mi\-nis\-terio de 
Econom\'{\i}a y Com\-pe\-ti\-ti\-vi\-dad
and the European regional development fund (FEDER)
under project number FIS2011-28738-C02-02,
by Junta de Andaluc\'{\i}a under project P07-FQM-02962,
and by Spanish Consolider-Ingenio 2010 (CPANCSD2007-00042).
J.E.G.R. and P.V.I. acknowledge support from the Spanish-French 
collaboration project AIC-D-2011-0676.

\end{document}